\documentclass[prc,aps,twocolumn,floatfix,superscriptaddress]{revtex4}                                                                                                                                                                                                                                                                                        
\usepackage{bm}
\usepackage{amsmath}
\usepackage{amsfonts}
\usepackage{amssymb}
\usepackage{graphicx}
\usepackage{color}
\usepackage{epstopdf}

\begin{document}

\title{New method of analytic continuation of elastic-scattering data to the negative-energy region and asymptotic normalization coefficients for $^{17}$O and $^{13}$C}

\author{L. D. Blokhintsev}
\affiliation{Skobeltsyn Institute of Nuclear Physics, Lomonosov Moscow State University, Moscow 119991, Russia}
\author{A. S. Kadyrov}
\affiliation{Curtin Institute for Computation and Department of Physics and Astronomy, Curtin University, GPO Box U1987, Perth, WA 6845, Australia}
\author{A. M. Mukhamedzhanov}
\affiliation{Cyclotron Institute, Texas A\&M University, College Station, Texas 77843, USA}
\author{D. A. Savin} 
\affiliation{Skobeltsyn Institute of Nuclear Physics, Lomonosov Moscow State University, Moscow 119991, Russia}

\begin{abstract}
A new method is proposed for extrapolation of elastic-scattering data to the negative-energy region for a short-range interaction. The method is based on the analytic approximation of the modulus-squared  of the partial-wave scattering amplitude. It is shown that the proposed method has an advantage over the traditional one based on continuation of the effective-range function. The new method has been applied to determine the asymptotic normalization coefficients for the $^{17}$O and $^{13}$C nuclei in the $n+^{16}$O and $n+^{12}$C 
channels, respectively. 
\end{abstract}

\maketitle

\section{Introduction} 
Neutron-induced processes and neutron transfer reactions play an important role in nuclear reactions, nuclear astrophysics, and applied physics. 
In recent years these reactions have  attracted a great interest due to their role in primordial nucleosynthesis of light elements \cite{RolfsRodney} and in inhomogeneous Big Bang models where $(n,\,\gamma)$ processes take part in reaction chains leading to the synthesis of heavy
elements \cite{Heil,Liu}.  While the elements lighter than iron are either created during the Big Bang or fusion reactions in stars, most of the elements heavier than iron are produced via neutron-induced reactions \cite{RolfsRodney}. Therefore, the knowledge of neutron-capture
cross sections for stable and unstable isotopes is essential. 
In many cases low-energy neutron radiative-capture reactions 
and neutron-transfer reactions populate loosely-bound states of final nuclei. To calculate the cross sections of such reactions one needs to know full information about the final bound states, in particular, their quantum numbers, binding energies, and asymptotic normalization coefficients (ANCs).

Using scattering data may give valuable information on ANCs, which, in contrast to binding energies, cannot be directly measured. The ANCs are fundamental nuclear characteristics that are important, for example, for evaluating cross sections of peripheral astrophysical nuclear reactions \cite{MukhTim,Xu,MukhTr,reviewpaper}.  One of the direct ways  of extracting ANCs from experimental data is the analytic continuation in the energy plane of the partial-wave elastic-scattering amplitudes, obtained by the phase-shift analysis,  to the pole corresponding to a bound state. Such a procedure, in contrast to the method of constructing optical potentials fitted to scattering data, allows one to circumvent an ambiguity problem associated with the existence of phase-equivalent potentials \cite{BlEr,BlOrSa}. 

The conventional procedure for such extrapolation is the analytic approximation of the experimental values of the  effective-range function 
(ERF) $K_l(E)$ with the subsequent continuation to the pole position ($l$ and $E$ are the orbital angular momentum and the relative kinetic energy of colliding particles, respectively). The ERF method has been successfully employed to determine the ANCs for bound (as well as resonant) nuclear states in a number of works (see, e.g. \cite{BKSSK,SpCaBa,IrOr} and references therein). 

In our previous works \cite{BKMS1,BKMS2,BKMS3} we investigated analytical continuation of scattering data for charged particles to the negative-energy region to obtain
information about ANCs. In the present paper, a new method is proposed for extrapolating data on elastic scattering of neutrons. When analyzing neutron scattering, in contrast to scattering of charged particles, one deals only with a short-range interaction. The method developed here makes use of the modulus-squared, denoted as $M_l(E)$, of the partial-wave scattering amplitude 
$f_l (E)$. Since $M_l(E)$ is a real analytic function of $E$ on the real positive semi-axis of $E$ including $E=0$, it can be analytically approximated by polynomials in $E$ for $E>0$ and then analytically continued to the bound state pole to obtain information on the ANC. 

Within an exactly solvable model, it is shown that the proposed method has an advantage over the traditional one based on the continuation of the ERF. Using the available data on phase shifts, two versions of the new method, along with the ERF method, have been applied to determine the ANCs for the 
$^{17}$O and $^{13}$C nuclei in the $n+^{16}$O and $n+^{12}$C 
channels, respectively.

Performing experiments on neutron elastic scattering is not an easy task. However, for heavier nuclei, where the Coulomb interaction significantly complicates extrapolation of the proton elastic-scattering phase shifts, progress in new experimental facilities  and methods can make  measurements of neutron elastic scattering a valuable technique to obtain information about neutron ANCs using the extrapolation method suggested in this paper. The method  provides faster convergence than the traditional one based on ERF. 
This is a significant advantage especially when experimental data have bigger uncertainties.
In addition, using the mirror symmetry one can determine the proton ANCs from the extracted neutron ANCs. 

The paper is organized as follows. In Sec. II, the theoretical backgrounds of the proposed method are outlined. Sections III and IV deal with the 
$n+^{16}$O and $n+^{12}$C systems,  
respectively.

 Throughout the paper we use the system of units in which $\hbar=c=1$.

\section{New method of analytic continuation of a partial-wave elastic-scattering amplitude}

Consider the partial-wave amplitude of elastic two-particle scattering $f_l (E)$ for a short-range interaction ($l$ is the orbital angular momentum, $E = k^2/2 \mu$ is the relative kinetic energy of colliding particles, $k$ is their relative momentum, $\mu$ is the reduced 
mass). Denote $E = E_+$ if $E>0$ and $E = E_-$ if $E<0$.

Suppose that in the system under consideration there is a bound state with energy $E=-\varepsilon = -\varkappa^2/2\mu <0$. 
For $E> 0$, we have
\begin{align}\label{f}
f_l(E_+)&=\frac{k^{2l}}{D_l(E_+)},\quad f^*_l(E_+)=\frac{k^{2l}}{D^*_l(E_+)}, \\
D_l(E_+)&=k^{2l+1}(\cot\delta_l-i).
\end{align}
Introduce a quantity $M_l(E)$ according to
\begin{align}\label{module}
 M_l(E_+)&\equiv|f_l(E_+)|^2
 =\frac{k^{4l}}{N_l(E_+)}, \\
  N_l(E_+)&=k^{4l+2}(\cot^2\delta_l+1).
\end{align}
Since  
\begin{align}\label{N}
N_l(E)&
=K_l^2(E)+k^{4l+2}, \\ 
K_l(E)&=k^{2l+1}\cot\delta_l,
\end{align}
and the effective-range function  $K_l(E)$ can be expanded in a series in $k^2$ near $k=0$,
the function $N_l(E)$ can also be expanded in a series in $k^2$ (or in $E$) near $E = 0$.
Therefore, one can approximate $N_l(E)$ with the expression                                                                                                                 
\begin{align}\label{N1}
N_l(E)=(E+\varepsilon)F_l(E),
\end{align}
where $F_l(E)$ is a polynomial or a rational function of $E$. The function $N_l(E)$ as given by Eq. (\ref{N1}) can be analytically continued to the domain  
$E<0$. The $E+\varepsilon$ factor provides the pole of the amplitude $f_l(E)$ at the energy corresponding to the bound state.
When $E\to -\varepsilon$
we have
\begin{align}\label{lim}                                                                     
\lim_{\substack{E\to -\varepsilon}}[(E+\varepsilon)M_l(E)]&=\lim_{\substack{E\to -\varepsilon}}
\left[(E+\varepsilon)\frac{k^{4l}}{(E+\varepsilon)F_l(E)}\right] \nonumber \\
&=\frac{\varkappa^{4l}}{F_l(-\varepsilon)}.
\end{align}

On the other hand, using the connection between the residue of $f_l(E)$ and the asymptotic normalization coefficient  $C_l$ (see, for example, \cite{BBD77,BKMS1}) and considering that as $E\to -\varepsilon$, $\cot\delta_l\to i$, we have
\begin{align}\label{lim1}
\lim_{\substack{E\to -\varepsilon}}[(E+\varepsilon)f_l(E)]=-\frac{1}{2\mu}C_l^2,  
\end{align}
\begin{align}\label{lim2}
\lim_{\substack{E\to -\varepsilon}}f_l^*(E)=\lim_{\substack{E\to -\varepsilon}}\frac{k^{2l}}{k^{2l+1}(\cot\delta_l+i)}=
-\frac{1}{2\varkappa}.
\end{align}
Combining (\ref{lim1}) and (\ref{lim2}), we get
\begin{align}\label{lim3}                                                                     
\lim_{\substack{E\to -\varepsilon}}[(E+\varepsilon)M_l(E)]&=\lim_{\substack{E\to -\varepsilon}}[(E+\varepsilon)f_l(E)f_l^*(E)] \nonumber \\
&=\frac{C_l^2}{4\mu\varkappa}.
\end{align}

Comparing (\ref {lim}) and (\ref {lim3}) gives the final result                                                                                                                                    
\begin{align}\label{C}                                                                     
C_l^2=\frac{4\mu\varkappa^{4l+1}}{F_l(-\varepsilon)}.
\end{align}
In this method, in contrast to the method based on the continuation of  the ERF $K_l(E)$, when defining the ANC $C_l$, there is no need to use the procedure of  differentiation, impairing the accuracy of the results.

Consider a slightly different version of the approximation of $M_l(E)$ for $E>0$:
\begin{align}\label{module1}
M_l(E)=|f_l(E)|^2=|e^{i\delta_l}\sin\delta_l/k|^2=\sin^2\delta_l/k^2.
\end{align}
Note that $\delta_l$ is an odd function of $k$ and $\sin^2\delta_l$ is an even function of $k$. Therefore, taking into account the threshold behavior of $\delta_l$, one can write:
\begin{align}\label{module2}
M_l(E)=\frac{k^{4l}}{E+\varepsilon}G_l(E),
\end{align}
where $G_l(E)$ is a polynomial or a rational function of $E$. From here, taking into account (\ref{lim3}), we obtain:
\begin{align}\label{lim4}
\lim_{\substack{E\to -\varepsilon}}[(E+\varepsilon)M_l(E)]=\varkappa^{4l}G_l(-\varepsilon)=\frac{C_l^2}{4\mu\varkappa}
\end{align}
and
\begin{align}\label{C1}
C_l^2=4\mu\varkappa^{4l+1}G_l(-\varepsilon).
\end{align}
The expression (\ref {C1}) differs from (\ref {C}) only by replacing $1/F_l(E)$ with $G_l(E)$.

Unfortunately, it is not clear how to generalize this method to include the Coulomb interaction since the renormalized Coulomb-nuclear partial-wave amplitude $\tilde f_l^*(E)$, unlike $\tilde f_l(E)$, has an essential singularity on the physical sheet of $E$ at $E=0$ 
 and is complex at $E<0$ \cite{BKMS1,BKMS2}.

\section{$n+^{16}$O system}

In this section, we consider the $n+^{16}$O system in the $J^{\pi}=1/2^+$ state, since only for this state data on the phase-shift analysis are available in the literature. By continuing these data to a point corresponding to the bound state energy $E = - \varepsilon_1 $ we  determine the ANC $C_{0}$ for the excited state of the nucleus $^{17}\mathrm {O}(1/2^+; 0.8707$ MeV) in the $n+^{16}$O (ground state) channel. Various continuation methods are compared: the continuation of the ERF $K_0 (E)$ and the continuation of the functions $F_0(E)$ and $G_0(E)$ introduced in Section II. Note that the  determination of the ANC for the mirror nucleus $^{17}$F in the $p+^{16}$O channel by extrapolating the elastic-scattering data was carried out in \cite{BKMS3}.

The following mass values are used in the calculations: $m_{^{17}\mathrm {O}}$ = 15830.501 MeV, $m_{^{16}\mathrm{O}}$ = 14895.079 MeV, and 
$m_n$ = 939.565 MeV.

\subsection{Theoretical  $n+^{16}$O phase shifts}

In this subsection, theoretical phase shifts $\delta_0$ calculated for the square-well potential from \cite{PhysRev.109.89} are used to compare different ways of continuing the scattering data to the negative-energy region. The parameters of the potential are: $V_0$ = 35.14 MeV, $R$ = 4.21 fm. This potential leads to two bound $s$-states, the lower of which is forbidden. The upper (allowed) state corresponds to the values of the binding energy $\varepsilon_1 = 3.59515$ MeV and ANC $C_{0} = 2.83896$ fm$^{- 1/2}$. Note that a more accurate experimental value of the binding energy is $\varepsilon_1 = 3.27227$ MeV.

This paper uses a more traditional deviation estimate 
based on the method of least squares
\cite{Wolberg}, 
\begin{align}\label{chi2}
\chi^2 =\frac{1}{N_p-N_f} \sum_{i=1}^{N_p} \left[ \frac{F(E_i) -f(E_i)}{\epsilon_i} \right]^2, 
\end{align}
where $N_p$ is the number of points, and $N_f$ is the number of parameters of the approximating function, $\epsilon_i$ is the error of the approximated function. 
Equation \eqref{chi2} takes into account the number of degrees of freedom and has several advantages over the definition used in the previous work \cite{BKMS3}.
This paper uses the approximation of continued functions by polynomials in energy $E$. For a polynomial of degree $N$, $N_p-N_f  = N_p-N-1$. 

For theoretical phase shifts, the errors of the approximated functions are assumed to be equal to each other (for simplicity, 
$\epsilon_i = 1$ for all $i$).
We start with the continuation of the ERF $K_0(E)$. While continuing $K_0(E)$ in all calculations in this work, a point corresponding to the energy of a bound state $E=-\varepsilon$ is added to points where phase shifts are known.

\begin{table}[htb]
\caption{ANC obtained by approximating ERF $K_0(E)$ for the $n+^{16}$O, $J^{\pi}=1/2^+$ state using a polynomial of degree~$N$.}
\begin{center}
\begin{tabular}{ccc}
\hline \hline
$N$ & $C_{0}$, fm$^{-1/2}$ & $\chi^2$ \\ 
\hline 
1 & -                               & 0.360438$\times 10^{-3}$ \\
2 & 2.06155 &  0.204316$\times 10^{-5}$ \\
3 & 5.33880 & 0.104794$\times 10^{-7}$  \\
4 & 2.73289  & 0.382649$\times 10^{-10}$  \\
5 & 3.08486 & 0.916381$\times 10^{-13}$  \\
6 & 2.74633 & 0.221163$\times 10^{-15}$  \\
7 & 2.88457 &  0.811685$\times 10^{-18}$ \\
8 & 2.81882 &  0.130191$\times 10^{-19}$ \\
9 & {2.84815} & {0.143192$\times 10^{-20}$}  \\
10 & 2.83505 & 0.666668$\times 10^{-20}$  \\
11 & - & 0.153103$\times 10^{-15}$  \\
\hline
exact & 2.83896 &  \\
\hline \hline
\end{tabular}
\end{center}
\label{table1}
\end{table}

The results of the continuation are presented 
in Table \ref {table1}. 
As one can see, 
for large degrees of the approximating polynomial, a breakdown occurs due to the excess of accuracy, and the approximation becomes very different from the approximated function. 
It can be seen that the visible breakdown occurs at $N = 11$. The best ANC value according to the  $\chi ^ 2$ criterion corresponds to $N = 9$ and is equal to 
$C_{0}$ = 2.84815 fm $^{-1/2}$, that is, the deviation from the exact value is about 0.3 \%. Dashes in the tables indicate the absence of a bound state with the correct theoretical energy.

\begin{figure}[htb]
\center{\includegraphics[width=0.9\columnwidth]{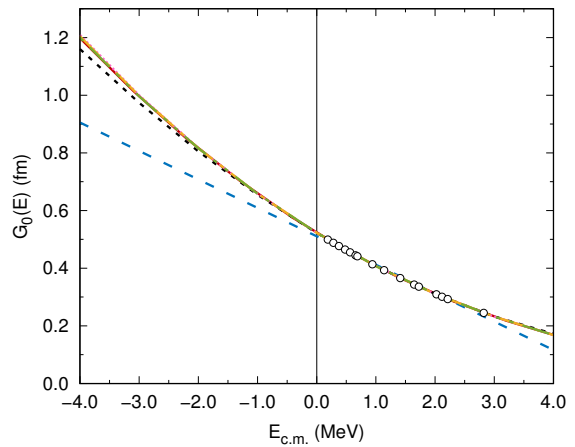}}
\caption {Function $G_0(E)$ for $n+^{16}$O,  $J^{\pi}=1/2^+$. 
Solid red line represents the results obtained from the theoretical phase shifts, 
long-dashed blue line - the 1st-order polynomial, 
short-dashed black line - the 2nd-order polynomial, 
dotted pink line - the 3rd-order polynomial, 
dash-dotted yellow line - the 4th-order polynomial, 
dash-double-dotted green line - the 5th-order polynomial. 
Starting from the 3rd-order polynomial, the results are indistinguishable from the solid red line.
}
\label{Figure1}
\end{figure}

We now consider the continuation of the function $G_0(E)$ introduced in Eq. (\ref{module2}). The results of the continuation are presented in Fig. \ref {Figure1} and in Table \ref {table2}. The best result is achieved again with $N=9$. 
As we can see the ANC at $N = 9$ reproduces the exact ANC to six significant digits.

\begin{table}[htb]
\caption{ANC obtained by approximating function $G_0(E)$ for the $n+^{16}$O, $J^{\pi}=1/2^+$ state using a polynomial of degree~$N$.}
\begin{center}
\begin{tabular}{ccc}
\hline \hline
$N$ & $C_{0}$, fm$^{-1/2}$ & $\chi^2$ \\ 
\hline 
1 &  2.50251 & 0.278856$\times 10^{-4}$\\
2 &  2.79892 & 0.204894$\times 10^{-7}$  \\
3 &  2.84871 & 0.255913$\times 10^{-11}$  \\
4 &  2.84452 & 0.285314$\times 10^{-13}$ \\
5 &  2.84042 & 0.270672$\times 10^{-16}$  \\
6 &  2.83926 & 0.197367$\times 10^{-19}$  \\
7 &  2.83902 & 0.743249$\times 10^{-23}$  \\
8 &  2.83897 & 0.217671$\times 10^{-26}$ \\
9 &  {2.83896} & {0.548333$\times 10^{-28}$}  \\
10 & 2.83897 & 0.668000$\times 10^{-28}$ \\
\hline
exact & 2.83896 &  \\
\hline \hline
\end{tabular}
\end{center}
\label{table2}
\end{table}

Finally, we consider the continuation of the function $F_0(E)$ introduced in Eq. (\ref{N1}). The results of the extrapolation are presented in Table \ref {table3}. The best result corresponds to $N=12$ and the relative error of the ANC at $N = 12$ with respect to the exact ANC is $1.5 \times 10^{-4}$.

\begin{table}[htb]
\caption{ANC obtained by approximating function $F_0(E)$ for the $n+^{16}$O, $J^{\pi}=1/2^+$ state using a polynomial of degree~$N$.}
\begin{center}
\begin{tabular}{ccc}
\hline \hline
$N$ & $C_{0}$, fm$^{-1/2}$ & $\chi^2$ \\ 
\hline 
1 & -  & 0.652168$\times 10^{-2}$ \\
2 & 1.73852  & 0.826223$\times 10^{-4}$ \\
3 & -  & 0.542466$\times 10^{-6}$ \\
4 & 2.22777  & 0.205841$\times 10^{-8}$ \\
5 & 3.48689 & 0.695829$\times 10^{-11}$ \\
6 & 2.61596 & 0.333772$\times 10^{-13}$ \\
7 & 2.98402  & 0.989157$\times 10^{-16}$ \\
8 & 2.77309  & 0.250929$\times 10^{-18}$ \\
9 & 2.87203  & 0.753178$\times 10^{-21}$ \\
10 &  2.82183 & 0.293600$\times 10^{-23}$ \\
11 & 2.84813  & 0.380000$\times 10^{-26}$ \\
12 & 2.83852 & 0.183333$\times 10^{-26}$ \\
13 &  {2.85509}  & {0.150000$\times 10^{-26}$} \\
14 & 2.62132  & 0.220000$\times 10^{-26}$ \\
\hline
exact & 2.83896 &  \\
\hline \hline
\end{tabular}
\end{center}
\label{table3}
\end{table}

Comparison of the data from Tables I-III reveals that the fastest convergence with increasing degree $N$ of the approximating  polynomial and the highest accuracy of the results for ANC $C_{0}$ occur in the case of approximation of the function $G_0(E)$. In fact, in this case a good level of convergence is achieved already at $N = 3$.

\subsection{Experimental $n+^{16}$O phase shifts}

In this subsection, we use 16 values of phase shifts $\delta_0$ from \cite{PhysRev.109.89, PhysRevC.2.124, PhysRev.162.890}, which correspond to the following neutron energy values $E_n$ in the laboratory system: $E_n$ = [0.20, 0.30, 0.40, 0.51, 0.60, 0.698, 0.73, 1.00, 1.21, 1.50, 1.75, 1.833, 2.15, 2.250, 2.353, 3.000] MeV.

For illustration, we also use the theoretical square-well potential with the parameters $V_0$ = 34.90941226 MeV, $R$ = 
4.191822098 fm. This potential is close to the potential used in subsection III A. For the upper (allowed) $s$-state of $^{17}$O, it leads to the correct experimental binding energy $\varepsilon_1 = 3.27227$ MeV and ANC $C_{0}$ = 2.6 fm$^{-1/2}$. As in subsection III A, we compare the results of the extrapolation of the functions $K_0(E)$, $G_0(E)$, and $F_0(E)$.

\begin{figure}[htb]
\center{\includegraphics[width=0.9\columnwidth]{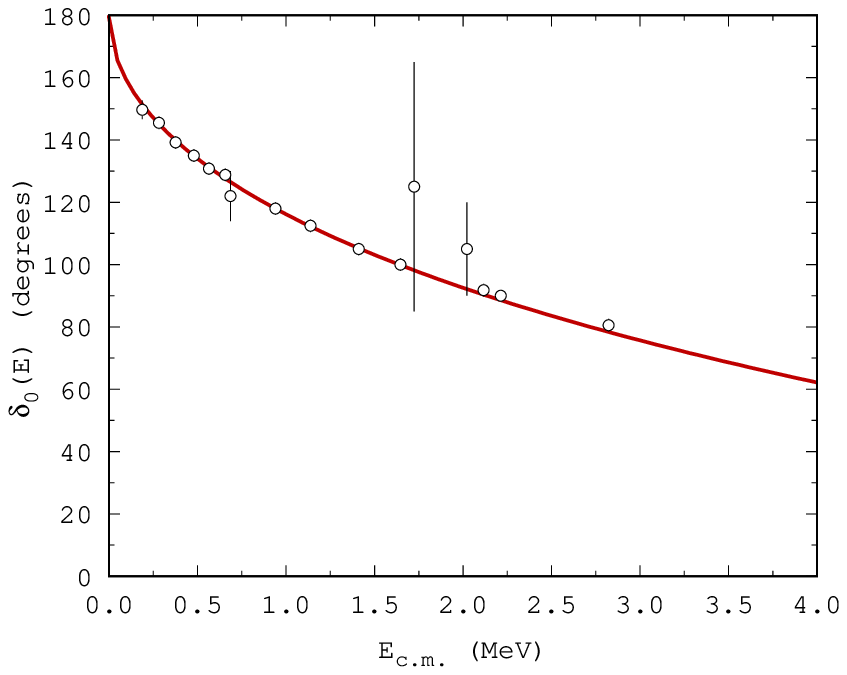}}
\caption {Experimental and theoretical phase shifts for $n+^{16}\mathrm{O}$,  $J^{\pi}=1/2^+$. The experimental points are from \cite{PhysRev.109.89, PhysRevC.2.124, PhysRev.162.890}. The theoretical results are obtained using the square-well potential described in the text.}
\label{Figure2}
\end{figure}

Experimental and theoretical phase shifts for the $n+\protect{^{16}{\rm O}}$ system in the  $J^{\pi}=1/2^+$ state are depicted in Fig. \ref{Figure2}. We see that the above potential describes the experimental data quite well.

\begin{figure}[htb]
\center{\includegraphics[width=0.9\columnwidth]{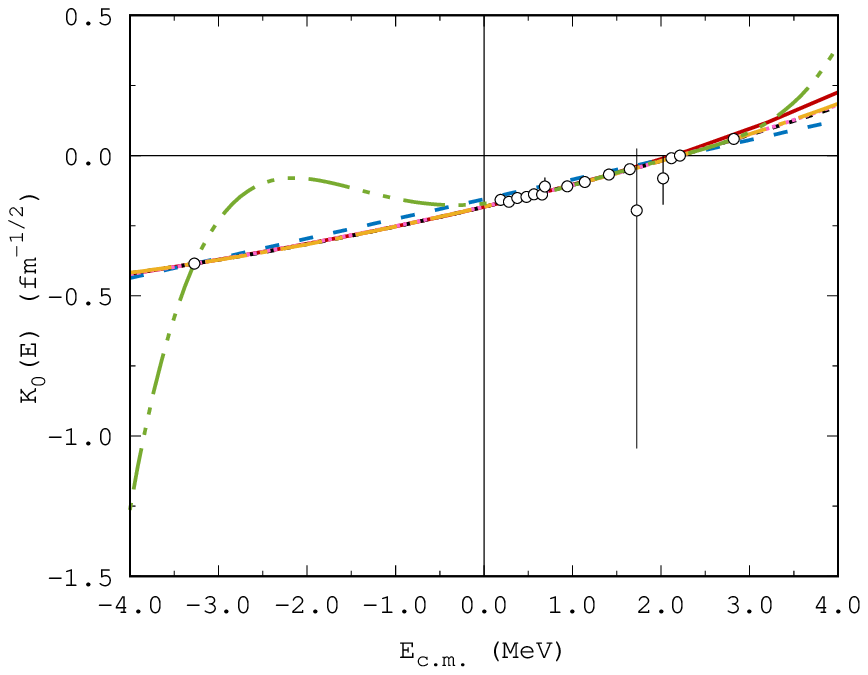}}
\caption {ERF for $n+^{16}$O, $J^{\pi}=1/2^+$. 
Solid red line represents the results obtained from theoretical phase shifts, 
long-dashed blue line - the 1st-order polynomial, 
short-dashed black line - the 2nd-order polynomial, 
dotted pink line - the 3rd-order polynomial, 
dash-dotted yellow line - the 4th-order polynomial, 
dash-double-dotted green line - the 5th-order polynomial. 
Starting from the 3rd-order polynomial, the results are indistinguishable.
Points represent the results obtained from the experimental phase shifts. 
}
\label{Figure3}
\end{figure}

\begin{figure}[htb]
\center{\includegraphics[width=0.9\columnwidth]{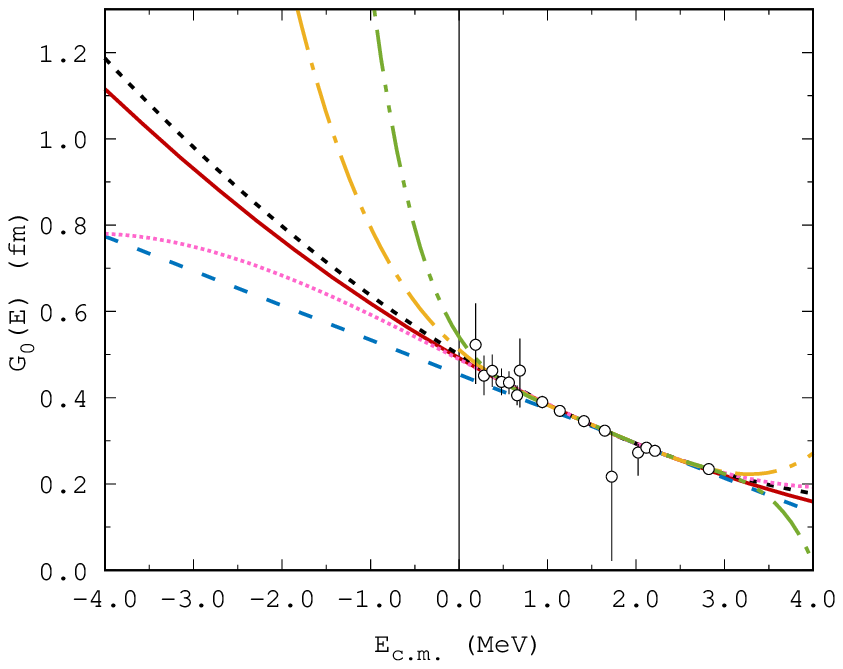}}
\caption {The same as in Fig. \ref{Figure3} but for function $G_0(E)$.}
\label{Figure4}
\end{figure}

\begin{figure}[htb]
\center{\includegraphics[width=0.9\columnwidth]{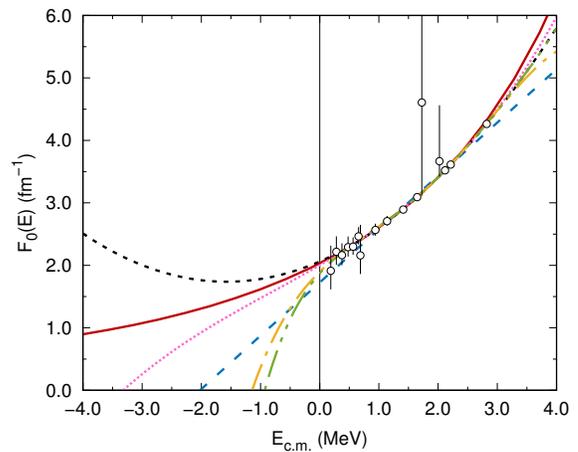}}
\caption {The same as in Fig. \ref{Figure3} but for function $F_0(E)$.}
\label{Figure5}
\end{figure}

The results of the extrapolation of EFR $K_0(E)$ are presented in Fig. \ref {Figure3}. As can be seen from this figure, for large degrees of the approximating polynomial, a breakdown occurs, and the approximation becomes very different from the approximated function. The best variant according to the $\chi^2$ criterion is $N = 2$ and leads to $C_{0}$ = 2.20716 fm$^{-1/2}$.
In case of continuing $G_0(E)$ (see Fig. \ref {Figure4}), the best ANC value by the $\chi^2$ criterion is $C_{0}$ = 2.67254
fm$^{-1/2}$, which, as in the case of EFR continuation, corresponds to $N = 2$.
When extrapolating the $F_0(E)$ function (Fig. \ref {Figure5}), again, the $N = 2$ variant is best by the $\chi^2$ criterion leading to  
$C_{0}$ = 1.80667 fm$^{-1/2}$. 

We see that different ways of continuing the experimental data lead to slightly different results for the ANC $C_{0}$. This  may be due to the low accuracy of the phase shift analysis used. The mean value of $C_{0}$, corresponding to the above three values, is 
$C_{0} = 2.23 \pm 0.30$ fm$^{-1/2}$.

\section{$n+^{12}$C system} 

This section discusses the $n+^{12}$C system in the $1/2^+$ state for which phase-shift data are available. By continuing the scattering data to a point corresponding to the experimental energy of the bound state $E = - \varepsilon_2 = 1.856557$  MeV, the ANC $C_{0}$ is determined  for the excited state of the nucleus $^{13} \mathrm {C} (1/2^+; 3.089$ MeV) in the channel $n+^{12}$C(ground state). As in Section III, the results obtained by extrapolating  functions $K_0 (E)$, $F_0(E)$, and $G_0(E)$ are compared.  The following mass values are used: 
$m_{^{13}\mathrm {C}}$ = 12109.481 MeV, $m_{^{12}\mathrm {C}}$ = 11174.862 MeV, and $m_n$ = 939.565 MeV.

\subsection{Theoretical phase shifts $n+^{12}$C}

In this subsection, the theoretical phase shifts $\delta_0$, calculated for the square-well potential  with the parameters $V_0$ = 
35.6753320221032 MeV and $R$ = 4.02818653449678 fm, are used to compare the effectiveness of various continuation methods. This potential leads to two bound $s$-states of $^{13}$C, the lower of which is forbidden. The upper (allowed) state corresponds to the experimental value of the binding energy $\varepsilon_2 = 1.856557$ MeV and ANC $C_{0} = 1.60$ fm$^{-1/2}$.

For theoretical phase-shift values, the errors of the approximated functions are assumed to be equal to each other (for simplicity, 
$\epsilon_i = 1$ for all $i$). 

\begin{figure}[htb]
\center{\includegraphics[width=0.9\columnwidth]{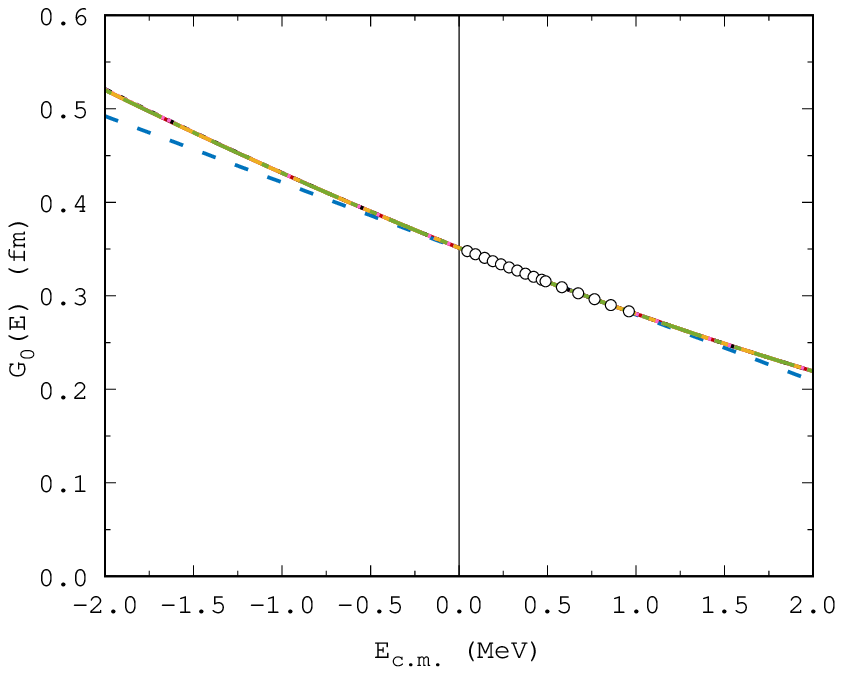}}
\caption {Function $G_0(E)$ for $n+^{12}$C,  $J^{\pi}=1/2^+$.  
Solid red line represents the results obtained from theoretical phase shifts, 
long-dashed blue line - the 1st-order polynomial, 
short-dashed black line - the 2nd-order polynomial, 
dotted pink line - the 3rd-order polynomial, 
dash-dotted yellow line - the 4th-order polynomial, 
dash-double-dotted green line - the 5th-order polynomial. 
Starting from the 2nd-order polynomial, the results are indistinguishable.
}
\label{Figure6}
\end{figure}

The results of the continuation of the functions $K_0(E)$, $G_0(E)$, and $F_0(E)$ are presented 
in Tables \ref{table4}, \ref{table5} and \ref{table6}. For all continuation versions, the best ANC values $C_{0}$ by the  $\chi^2$ criterion are close to the exact result. Comparing Tables IV-VI we conclude that, as in the case of the $n+^{16}$O system, the fastest convergence with increasing degree $N$ of the approximating polynomial and the highest accuracy of the results for ANC $C_{0}$ takes place in the case of extrapolating the function $G_0(E)$. The results of the continuation of the function $G_0(E)$ are shown in Fig. \ref{Figure6}.

\begin{table}[htb]
\caption{ANC obtained by approximating ERF $K_0(E)$ for the $n+^{12}$C, $J^{\pi}=1/2^+$ state using a polynomial of degree~$N$.}
\begin{center}
\begin{tabular}{ccc}
\hline \hline 
$N$ & $C_{0}$, fm$^{-1/2}$ & $\chi^2$ \\ 
\hline 
1  &  2.28097  & 0.867724$\times 10^{-5}$  \\
2  &  1.52384  & 0.448646$\times 10^{-8}$ \\
3  &  {1.60788}  & {0.434737$\times 10^{-10}$} \\
4  &  1.11353  & 0.411513$\times 10^{-8}$ \\
5  &  0.157698 & 0.431582$\times 10^{-6}$ \\
\hline
exact & 1.60 &  \\
\hline \hline
\end{tabular}
\end{center}
\label{table4}
\end{table}

\begin{table}[htb]
\caption{ANC obtained by approximating function $G_0(E)$ for the $n+^{12}$C, $J^{\pi}=1/2^+$ state using a polynomial of degree~$N$.}
\begin{center}
\begin{tabular}{ccc}
\hline \hline 
$N$ & $C_{0}$, fm$^{-1/2}$ & $\chi^2$ \\ 
\hline 
1 & 1.56036 & 0.125628$\times 10^{-6}$ \\
2 & 1.60147 & 0.131839$\times 10^{-12}$ \\
3 & 1.60109 & 0.816387$\times 10^{-14}$ \\
4 & 1.60018 &  0.243129$\times 10^{-17}$\\
5 & 1.60002 & 0.330466$\times 10^{-21}$ \\
6 & 1.60000 & 0.263367$\times 10^{-25}$ \\
7 & {1.60000} & {0.513375$\times 10^{-27}$} \\
8 & 1.60000 & 0.132814$\times 10^{-26}$ \\
9 & 1.60002 & 0.213933$\times 10^{-26}$ \\
10 & 1.59955 & 0.286980$\times 10^{-26}$ \\
\hline
exact & 1.60 &  \\
\hline \hline 
\end{tabular}
\end{center}
\label{table5}
\end{table}

\begin{table}[htb]
\caption{ANC obtained by approximating function $F_0(E)$ for the $n+^{12}$C, $J^{\pi}=1/2^+$ state using a polynomial of degree~$N$.}
\begin{center}
\begin{tabular}{ccc}
\hline \hline 
$N$ & $C_{0}$, fm$^{-1/2}$ & $\chi^2$ \\ 
\hline 
1 & 1.83289 & 0.708967$\times 10^{-4}$ \\
2 & 1.54553 & 0.700486$\times 10^{-7}$ \\
3 & 1.61637 & 0.616769$\times 10^{-10}$ \\
4 & 1.59590 & 0.386920$\times 10^{-13}$ \\
5 & 1.60102 & 0.238136$\times 10^{-16}$ \\
6 & 1.59976 & 0.111939$\times 10^{-19}$ \\
7 & 1.60006 & 0.548256$\times 10^{-23}$ \\
8 & {1.59999} & {0.119700$\times 10^{-24}$} \\
9 & 1.59990 & 0.213267$\times 10^{-24}$ \\
10 & 1.60057 & 0.298720$\times 10^{-24}$ \\        
\hline
exact & 1.60 &  \\
\hline \hline
\end{tabular}
\end{center}
\label{table6}
\end{table}

\subsection{Experimental $n+^{12}$C phase shifts}

We use 16 neutron-energy points (laboratory system) from \cite{Dub}: $E_n$ = [0.050, 0.100, 0.157, 0.207, 0.257, 0.307, 0.357, 0.407, 0.457, 0.507, 0.530, 0.630, 0.730, 0.830, 0.930, 1.040] MeV.

Phase-shift errors are assumed to be $\pm 1^\circ$. Note that increasing errors to $\pm 2^\circ$ only leads to negligible changes in the results.

\begin{figure}[htb]
\center{\includegraphics[width=0.9\columnwidth]{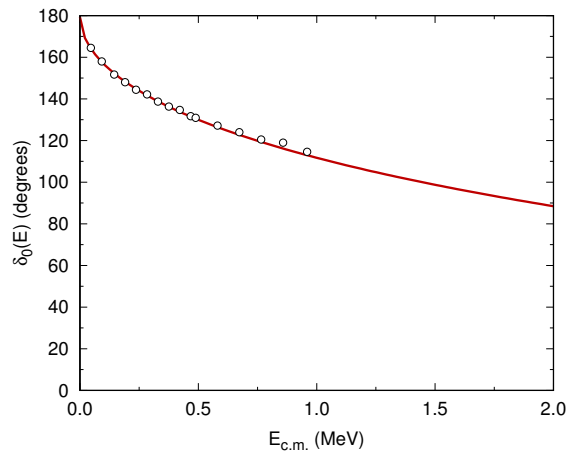}}
\caption {Experimental and theoretical phase shifts for $n+^{12}$C,  $J^{\pi}=1/2^+$.
The experimental points are from \cite{Dub}. The theoretical results are obtained using the square-well potential described in the text.}
\label{Figure7}
\end{figure}

Experimental and theoretical phase shifts for the $n+\protect{^{12}{\rm C}}$ system in the  $J^{\pi}=1/2^+$ state are depicted in Fig. \ref{Figure7}. Theoretical phase shifts are calculated using the potential described in Subsection IV A. As in the case of the 
$n+\protect{^{16}{\rm O}}$ system, there is good agreement between theory and experiment.

\begin{figure}[htb]
\center{\includegraphics[width=0.9\columnwidth]{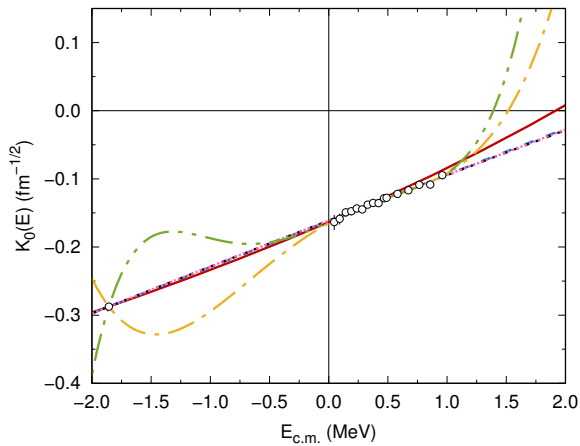}}
\caption {ERF for $n+^{12}$C,  $J^{\pi}=1/2^+$.   
Solid red line represents the results obtained from theoretical phase shifts, 
long-dashed blue line - the 1st-order polynomial, 
short-dashed black line - the 2nd-order polynomial, 
dotted pink line - the 3rd-order polynomial, 
dash-dotted yellow line - the 4th-order polynomial, 
dash-double-dotted green line - the 5th-order polynomial. 
Points represent the results obtained from the experimental phase shifts. 
}
\label{Figure8}
\end{figure}

\begin{figure}[htb]
\center{\includegraphics[width=0.9\columnwidth]{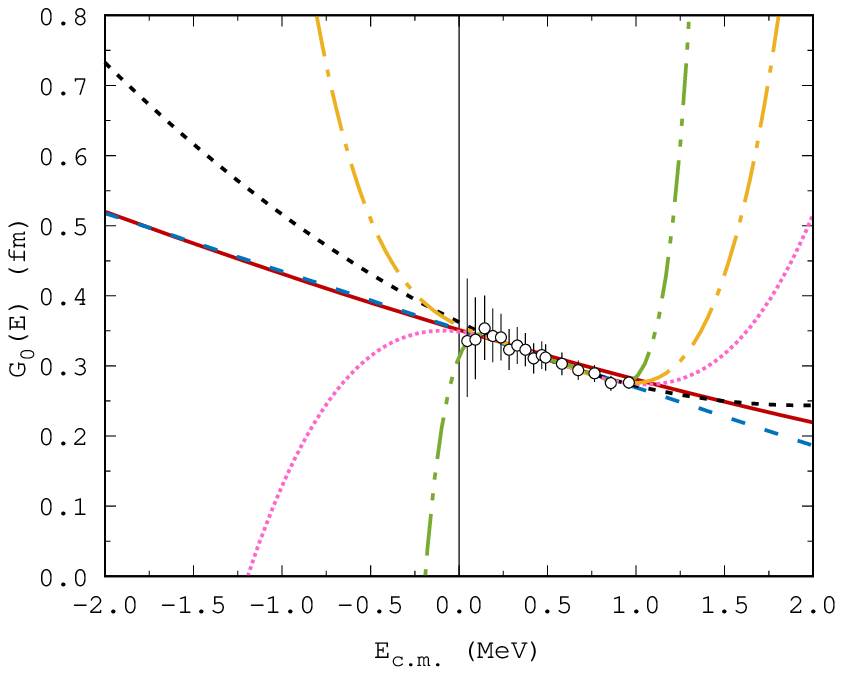}}
\caption {The same as in Fig. \ref{Figure8} but for function $G_0(E)$.}
\label{Figure9}
\end{figure}

\begin{figure}[htb]
\center{\includegraphics[width=0.9\columnwidth]{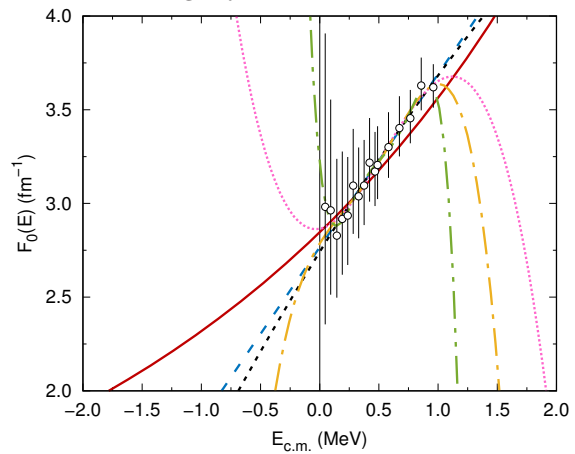}}
\caption {The same as in Fig. \ref{Figure8} but for function $F_0(E)$.}
\label{Figure10}
\end{figure}

The results of the continuing the ERF $K_0(E)$ are presented in Fig. \ref {Figure8}. The best ANC value by the $\chi^2$ criterion   corresponds to $N = 1$ and is equal to  $C_{0}$ = 2.14638 fm$^{-1/2}$.
With the continuation of the function $G_0(E)$ (Fig. \ref {Figure9}), the best ANC value is $C_{0}$ = 
1.87563 fm$^{-1/2}$, corresponding to $N = 2$.
Extrapolating the function $F_0(E)$ (Fig. \ref {Figure10}) leads to the best value of $C_{0}$ = 2.19107 fm$^{-1/2}$, corresponding to $N = 1$.
The mean value of $C_{0}$, corresponding to the above three values, is $C_{0} = 2.07 \pm 0.13$ fm$^{-1/2}$.

\section{Conclusions}

In the present paper, we proposed  a new method of extrapolating elastic scattering data to the negative energy region for a short-range interaction. It is shown that the proposed method has an advantage over the traditional method of continuing the effective-range function. Using the available phase-shift data, two versions of the new method, as well as the ERF method, have been applied to determine the ANCs for the excited $s$ states of $^{17}$O and $^{13}$C nuclei in the $n+^{16}$O and $n+^{12}$C channels, respectively. Due to the low accuracy of the phase-shift analysis used different ways of continuing the experimental data lead to slightly different results for the 
ANCs. The mean values of the ANCs obtained with all different methods used in this paper are 
$2.23 \pm 0.30$ fm$^{-1/2}$ for $^{17}$O and 
$2.07 \pm 0.13$ fm$^{-1/2}$ for $^{13}$C. For comparison, the ANC values obtained from the analysis of data on radiative neutron capture are 
3.01 fm$^{-1/2}$ for $^{17}$O and 
1.61 fm$^{-1/2}$ for $^{13}$ C \cite{Huang}. These  results are based on the assumption of the peripheral character of the $s$-wave radiative capture which is not justified. Therefore, the accuracy of these ANC values is difficult to estimate.
On the other hand, the method proposed in this work is equally suitable for extrapolation of elastic scattering data for any~$l$.

\section*{Acknowledgements}

This work was supported by the Russian Science Foundation
Grant No. 16-12-10048 (L.D.B.) and the Russian Foundation
for Basic Research Grant No. 18-02-00014 (D.A.S.).
A.S.K. acknowledges a support from the Australian Research
Council and the U.S. NSF Grant No.
PHY-1415656. A.M.M. acknowledges support from the U.S. DOE
Grant No. DE-FG02-93ER40773, the U.S. NSF Grant No.
PHY-1415656, and the NNSA Grant No. DE-NA0003841.

\end{document}